\documentclass[aps, reprint, floatfix]{revtex4-1}

\usepackage{amsmath,amssymb,graphicx}
\usepackage{enumitem}
\usepackage{amsbsy}
\usepackage{latexsym}
\usepackage{color}
\usepackage{graphicx}
\usepackage{psfrag}
\usepackage[normalem]{ulem}
\usepackage{bm}
\usepackage{lipsum}
\usepackage{color}
\usepackage{tikz}
\usepackage{dcolumn}
\usepackage{multirow}
\usepackage{graphicx, amssymb, verbatim, amsmath, adjustbox, siunitx, color, lmodern, float, eucal, soul, makecell, color, times, caption}

\usepackage{natbib,url}
\usepackage[colorlinks, allcolors = blue, bookmarksopen,bookmarksnumbered]{hyperref}
\usepackage{cleveref}
\usepackage[T1]{fontenc}
\usepackage{subfiles}

\begin{document}

\title
{Suppression of Active Super-Diffusion: Impact of String Defects and Canted Multi-Domains}
\author{Ritik Rajak}
\email{srz228579@sire.iitd.ac.in}
\affiliation{School of Interdisciplinary Research, Indian Institute of Technology, Hauz Khas, New Delhi 110016, India.}

\author{Manish Agarwal }
\email{zmanish@cc.iitd.ac.in}
\affiliation{Computer Services Centre, Indian Institute of Technology, Hauz Khas, New Delhi 110016, India.}

\author{Sanjay Puri}
\email{purijnu@gmail.com}
\affiliation{School of Physical Sciences, Jawaharlal Nehru University, New Delhi 110067, India.}

\author{Varsha Banerjee}
\email{varsha@physics.iitd.ac.in}
\affiliation{Department of Physics, Indian Institute of Technology, Hauz Khas, New Delhi 110016, India.}

\begin{abstract}

We investigate the transport dynamics of an active Brownian particle (ABP) traversing a complex, non-Newtonian liquid crystal (LC) matrix. Employing the Generalized Lebwohl–Lasher (GLL) model, we systematically vary higher-order orientational interactions to stabilize three distinct host environments: isotropic, uniform nematic, and structurally frustrated canted phases. Modeling the coupled system via off-lattice over-damped Langevin dynamics, the resulting trajectories are characterized by evaluating their step-size distributions (SSDs), mean-square displacements (MSDs), and Hurst exponents. 
In the uniform nematic phase, the anisotropic matrix elastically channels the ABP, producing a left-skewed exponential SSD and persistent ballistic motion parallel to the director $\hat{\mathbf{n}}$. Similarly, transverse transport obeys a Rayleigh distribution and acquires a prominent $t \ln t$ super-diffusive correction—an explicit signature of the particle coupling to the host's gapless transverse Goldstone modes, as predicted by Toner et al.~\cite{toner2016following_ref02}. Crucially, we reveal that this active super-diffusion is systematically suppressed when the long-range Goldstone fluctuations are disrupted by topological defects. This breakdown manifests both macroscopically within the fractured, multi-domain canted phase due to a structural mass gap, and locally in the unfrustrated nematic phase through scattering by vortex disclination lines. Consequently, while the local SSDs qualitatively mirror the ideal nematic state, the transverse $t \ln t$ scaling vanishes in the presence of these structural constraints. Our findings demonstrate that tuning the background defect architecture of a complex fluid can fundamentally alter the transport universality class of active matter, offering a novel paradigm for controlling microscopic mobility.
\end{abstract}

\maketitle

\section{Introduction}
Microorganisms are essential to our ecosystem and are profoundly shaped by their environment. Physical and chemical factors within the environment determine their ability to survive, grow, thrive, or enter hibernation. In addition, it serves as a key driver of their evolution, allowing some to adapt to extreme conditions. The properties of the medium in which the microorganisms reside significantly influence their movement and their trajectories, which have far-reaching consequences. For example, these movements affect dispersal and migration, which in turn impact ecological, environmental, and biological processes. The effects ripple beyond individual organisms, shaping entire ecosystems and global cycles. Microorganisms are also central to biogeochemical cycles, and their movements affect the rates and locations of these vital processes. In addition, the nature of their pathways is crucial for the formation, development, and dispersal of biofilms, which are structured microbial communities attached to surfaces. The spread of infectious diseases is closely related to the movements and transmission pathways of pathogenic microorganisms. Given the complexity of their natural habitats, the dynamics of microorganisms in these environments is not well understood.

A representative theoretical framework for understanding the behavior and motion of microorganisms is provided by active Brownian particles (ABPs) \cite{schimansky1995structure,romanczuk2012active, bechinger2016active}. They model their self-propelled motion in dissipative environments and represent a non-equilibrium generalization of classical Brownian particles. Much of the earlier work in this area 
has focused on ABPs in Newtonian fluids, which are characterized by constant viscosity and predictable uniform flow profiles. The relatively simple rheology of Newtonian fluids allows for tractable mathematical modeling of particle dynamics.
However, the natural environments in which microorganisms reside are often highly viscous and anisotropic, diverging significantly from Newtonian assumptions \cite{qiu2014swimming, shen2011undulatory, li2021microswimming, turiv2020topology, morales2019liquid, viney1993liquid}. Some examples include Streptococcus bacteria that navigate through blood, sperm cells that move through cervical mucus \cite{suarez2006sperm}, viruses that swim through nasal mucus, and E. coli that swims in polymeric solutions \cite{martinez2014flagellated, zottl2019enhanced, liu2021viscoelastic}. These complex viscoelastic media can strongly influence microbial locomotion by altering flagellar motion, thereby affecting both the speed and trajectory of swimming.
Although Newtonian models provide useful approximations, they fall short of describing the complex rheology of many biological fluids \cite{shen2011undulatory}. In such cases, a more accurate description involves non-Newtonian fluids, whose viscosities vary with applied stress or strain rate. These media can exhibit shear thinning, shear thickening, viscoelasticity, and other behaviors that significantly affect active particle dynamics \cite{goral2022frustrated}.

A particularly intriguing class of non-Newtonian fluids is liquid crystals (LCs) \cite{stephen1974physics, de1993physics, priestly2012introduction, andrienko2018introduction}. They are mesophases that are intermediate between liquids and solids, exhibiting both fluidity and long-range molecular order. These features make LCs highly responsive to external stimuli. Beyond their technological applications in display devices, LCs also occur naturally in systems such as cell membranes.  Among various types, nematic LCs are the simplest and most widely studied. Their rod-like or disk-like components align along a common axis called the director, imparting anisotropy to optical, elastic, and dielectric properties. Importantly, nematic LCs display non-Newtonian flow behavior: under shear, the molecules reorient along streamlines, giving rise to non-linear rheology. This coupling between flow and molecular alignment is central to the hydrodynamics of LCs and has significant implications for the motion of active particles and microorganisms in them.

Motility is a fundamental strategy for microbial survival, and understanding how it is modulated by structured environments such as LCs is crucial. In recent years, living liquid crystals (LLCs) have been explored - an amalgam of motile microorganisms, such as bacteria, suspended in biocompatible LC environments \cite{zottl2019enhanced, peng2016command,lavrentovich2016active, genkin2017topological, aranson2018harnessing,sokolov2019emergence, vats2025symbiotic, vats2025ferronematics}.
A unique feature of these systems is that the LC director field can be externally controlled or patterned, allowing for precise manipulation of active motion. This control enables diverse functionalities: circular director fields can confine active particles \cite{sokolov2019emergence}; uniform nematic alignment can direct propulsion for cargo transport; and splay–bend distortions near topological defects can generate autonomous motion \cite{peng2016command,genkin2017topological}. Consequently, LLCs have emerged as promising platforms for next-generation microfluidic technologies that operate without pumps or pressure gradients, as well as synthetic systems that mimic cellular motion and targeted drug-delivery applications \cite{kaiser2014transport, sokolov2015individual}. Furthermore, many biological assemblies such as densely packed DNA in bacteriophages, the tobacco mosaic virus, cellular membranes, and epithelial tissues exhibit liquid-crystalline order. 
Such naturally occurring liquid crystalline organizations also underscore the biological relevance of studying active motion in these environments.

An important piece of work in recent years is by Toner et al. \cite{toner2016following_ref02, toner2018_smectic} where they focus on the interplay between the self-propelled motion of the ABP in a nematic host. Because microorganisms operate in low Reynolds number regime, they adopted an over-damped Langevin description in which inertia is negligible and the instantaneous velocity of the particle reflects a balance among self-propulsion, medium interactions, and thermal noise. The LC environment was modeled using the Lebwohl-Lasher (LL) model, which captures the phase transition between the isotropic and nematic phases \cite{lebwohl1972nematic}. Toner et al. establish that the anisotropic nematic order qualitatively modifies active particle transport, giving rise to strongly direction-dependent dynamics. As time ($t$) evolves the motion along the nematic director is ballistic (MSD $\propto t^2$), while the transverse motion is super-diffusive with a $t \ln t$ scaling of the MSD \cite{toner2016following_ref02}. They showed that logarithmic corrections in the transverse direction are a direct consequence of Goldstone modes (massless excitations) inherent to the nematic phase that spontaneously breaks the continuous rotational symmetry of an isotropic fluid.
According to the Goldstone theorem, breaking of continuous symmetry must give rise to collective modes without gaps (massless). Unlike in standard liquids, where fluctuations are short-ranged, the Goldstone modes in nematic LCs lead to long-range correlations in the director field. The massive fluctuations are restricted to the transverse (perpendicular) direction relative to the global director ${\bf n}$, as changing the orientation longitudinally would require changing the fixed length of the nematogen.
Standard ABPs such a bacteria in water exhibit ballistic motion at short times (MSD $\propto t^2$) that crosses over to diffusive motion (MSD $\propto t$) at long times due to random thermal rotations. The presence of the $t \ln t$ term implies that the ABP never decays into a standard random walk, but picks up a diffusion coefficient that continuously grows with time ($D(t) \sim \ln t$) due to coupling with the long-range Goldstone modes. The latter provide an additional kick, making the ABP move faster than the standard diffusive process in the transverse direction. 

The above observations underscore how profoundly the underlying topology of liquid crystal textures dictates the directional transport characteristics of an ABP. Motivated, we investigate the dynamics of an active Brownian particle (ABP) in a LC medium that is represented by the generalized Lebwohl–Lasher (GLL) model \cite{zhang1993phase, chiccoli1997phase, Birdi_2020_ref03}. Compared with the standard LL formulation, the GLL model provides an accurate representation of interactions between rod-like mesogens and reproduces experimentally observed phase behavior. In particular, it captures complex LC textures, such as the canted phase,
in addition to the conventional nematic and isotropic phases. In the nematic phase, the molecular orientations are largely aligned along the director, whereas in the canted phase, neighboring molecules adopt characteristic angular offsets, producing a locally distorted orientational field. The anisotropic and non-Newtonian character of the LC medium is thus encoded in the evolving mesogen orientations, which themselves obey over-damped Langevin dynamics. By combining the stochastic motion of ABPs with the rich structural diversity of the LC host, our model enables a detailed exploration of how anisotropic environments influence active transport at the microscale. To quantify these effects, we analyze several key dynamical measures: (i) Hurst exponent ($H$) and fractal dimension ($d_f=d-H$) \cite{paul1997addison}; (ii) step-size distribution (SSD); (iii) mean-square displacement (MSD); and (iv) diffusion characteristics (e.g. ballistic, anomalous, sub- or super-diffusive regimes). Each of these metrics provides comprehensive insights into the particle's transport dynamics.

This study explicitly uncovers how distinct LC topologies govern active particle kinetics. Our central finding is that - unlike the unfrustrated nematic medium where long-range Goldstone fluctuations accelerate an ABP into a classic super-diffusive ($t \ln t$) regime - transport within the canted phase reveals a complete reversal of this paradigm. We demonstrate that higher-order microscopic interactions segment the canted morphology into an assembly of micro-domains bounded by sharp, interfacial defects. This structural compartmentalisation introduces a finite mass gap that suppresses long-wavelength orientational fluctuations, eliminating active superdiffusion. Crucially, we show that the presence of vortex strings - the topological defects of the uniform nematic medium - similarly disrupts these gapless Goldstone modes. Together, these observations prove that background defect architecture can completely flip the transport universality class of active matter, establishing a novel topological framework to control microscopic mobility.

The remainder of this paper is organized as follows. Sec.~\ref{Theoretical Framework} introduces the over-damped Langevin framework for the coupled ABP–GLL system. Sec.~\ref{Results} details the computational setup and results, and Sec.~\ref{Summary and Discussion} concludes with a summary and discussion. For convenience, we have provided Tables~\ref{Table A.1: model symbol list}--\ref{Table A.3: Symbols with values} that summarize the model parameters, scaled parameters and their numerical values in the Appendix. 

\section{Theoretical Framework}
\label{Theoretical Framework}

To describe the motion of the ABP in an LC medium, we first discuss the appropriate models that describe these systems. We begin with a description of the LC medium and ABP. Then we will describe the framework for the coupled system, which is governed by the competition between the co-alignment of the local nematic director and the velocity of the ABP, and the dissipative effects in the nematic medium. 

\subsection{Generalized Lebwohl-Lasher (GLL) Model for LCs}
At high temperatures, NLCs exist in an isotropic phase, characterized by randomly oriented rod-like molecules (nematogens) that possess both translational and rotational symmetry. Upon cooling, these nematogens tend to align statistically parallel along an arbitrary direction, forming the nematic phase, which exhibits a long-range orientational order extending over thousands of molecules. In the case of a uniaxial nematic, this orientation order is described by a sign-invariant unit vector known as the nematic director, denoted by $\hat{\mathbf{n}}$. A theoretical framework for understanding the isotropic-nematic phase transition was first developed by Maier and Saupe (MS), who employed an approximate molecular field theory to derive the pseudo-potential between two rod-like nematogens. Although the MS theory correctly predicted a first-order transition, its quantitative predictions for the orientational order parameter were inaccurate. To improve upon this, Humphries et al. extended the MS theory by expanding the pseudo-potential in a complete set of Legendre polynomials, $P_l(z)$. Due to the inversion symmetry of rod-like molecules, only even-order polynomials ($P_2, P_4, \ldots$) contribute, and the corresponding coefficients decrease rapidly with increasing $l$. The original MS theory is recovered when the summation is truncated at $l=2$.
\begin{figure*}[hbt]
    \centering
    \includegraphics[width=1\textwidth]{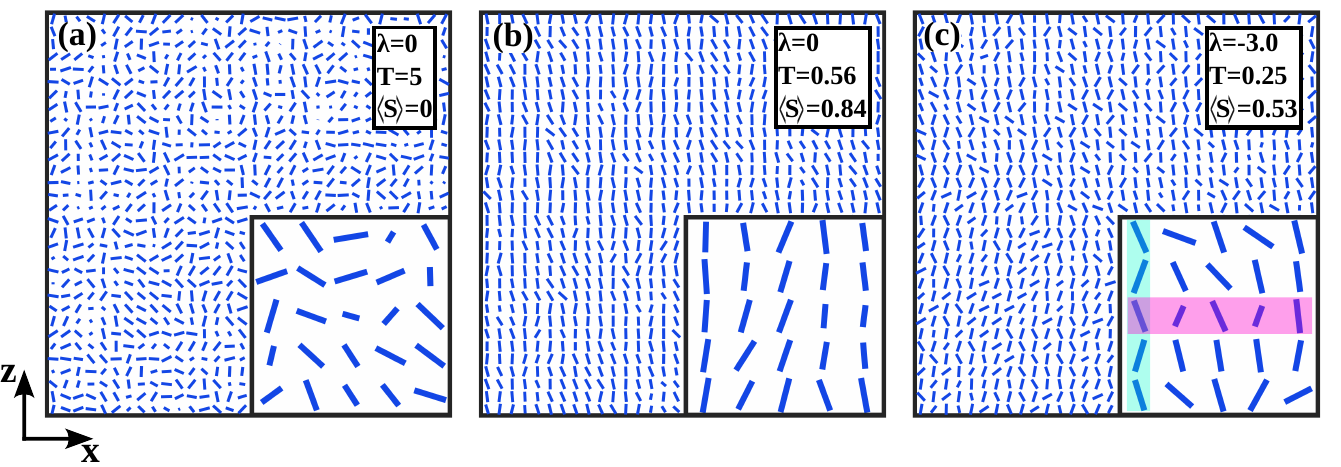}
    \captionsetup{justification=raggedright,
    singlelinecheck=false}
    \caption{Typical equilibrated morphology slices at $t=100$. The depiction is a  $32\times32$ cross-section of $xz$-plane at $y=0$ extracted from the $64^3$ system in (a) isotropic phase, (b) nematic phase, and (c) canted phase. The insets show a $5\times 5$ array of LC molecules for clarity of their orientation. In particular, distinct zigzag arrangements in the canted phase have been highlighted. The ensemble averaged values of the orientational order parameter $\langle \mathcal{S} \rangle$ for each phase have also been specified.}
    \label{phases}
\end{figure*}
The simplest three-dimensional ($d=3$) lattice version of MS theory is the LL model \cite{lebwohl1972nematic}. Each lattice site $i$ hosts a rod-like molecule with orientation of the long axis defined by a three-component ($n=3$) unit vector ${\hat{\mathbf{n}}_i=\left(\sin\theta_i\cos\phi_i, \sin\theta_i\sin\phi_i, \cos\theta_i\right)}$ with inclination $\theta_i \in [0,\pi]$ and azimuth $\phi_i \in [0,2\pi)$. The Hamiltonian for the LL model is given by:
\begin{equation}
\label{equation_1}
H = -\epsilon\sum_{\langle ij \rangle}P_2\left(\cos\theta_{ij}\right),
\end{equation}
where $\epsilon$ is the strength of the interactions between the nearest neighbors (nn), $\theta_{ij}$ is the angle between the nn nematic rods (${\cos\theta_{ij}=\hat{\mathbf{n}}_i\cdot\hat{\mathbf{n}}_j}$) and $P_2(z)=\left(3z^2-1\right)/2$. In subsequent discussions, we set $\epsilon=1$ for convenience.  The orientational order parameter is given by:
\begin{equation}
\label{equation_2}
\mathcal{S} = \langle P_2\left(\cos\theta_{i}\right)\rangle = \left\langle \frac{3\cos^2\theta_{i}-1}{2}\right\rangle.
\end{equation}
In the above expression, $\cos\theta_i = \hat{\mathbf{n}}_i\cdot\hat{\mathbf{n}}_j$ and the angular brackets $\langle\cdot\cdot\cdot\rangle$ imply an ensemble average. The I phase corresponds to $\mathcal{S}=0$ while a fully aligned N phase has $\mathcal{S}=1$. The defects correspond to regions of low order or $\mathcal{S}\simeq 0$. 
Introducing dimensionless temperature $T^{*} = k_BT/\epsilon$, the LL model exhibits a {\it weak} first-order transition at $T^{*}\simeq 1.1$. (We drop the star in the further description.) 
However, it inherits the quantitative discrepancies of MS theory and does not capture other orientationally ordered phases that are observed in nematic LCs. Nevertheless, the LL model has been widely used in the literature to understand the nematic phase because of its simplicity, analytical tractability, and computational ease. 

The inclusion of higher-order Legendre polynomials makes the potential more ``peaked'' around certain relative orientations of neighboring nematogens. Studies reveal that their inclusion strengthens the first-order nature of the isotropic-nematic phase transition and produces improved agreement with experiments. Scattering experiments indicate that only terms up to $P_4$ in the expansion of the pseudo-potential are statistically significant. Consequently, a reasonable approximation representing the interaction between nn nematogens is provided by the GLL \cite{zhang1993phase, chiccoli1997phase}:
\begin{equation}
\label{equation_3}
H = -\epsilon\sum_{\langle ij \rangle}\big[P_2\left(\cos\theta_{ij}\right)+\lambda P_4\left(\cos\theta_{ij}\right)\big], 
\end{equation}
\begin{table}[H]
    \begin{ruledtabular}
        \begin{tabular}{ccccc}
             $\boldsymbol{\lambda}$ &$0$ &${-1.0}$  & ${-2.0}$  & ${-3.0}$ \\ \hline
             \makecell{ $ \boldsymbol{\theta_c}$}&$0$  & \makecell{${39.23^o}$} & \makecell{${44.18}^o$} & \makecell{${45.82}^o$}  \\ 
             $\boldsymbol{T^*_c}$ & $ 1.12$  & $0.80$  & $0.61$ & $0.50$\\ \hline 
        \end{tabular}
    \end{ruledtabular}
    \captionsetup{justification=raggedright}
    \caption{ Variation of the canted angle $\theta_c$ and the corresponding critical temperature (in reduced units) as a function of $\lambda$. The data for $T^*_c$ has been extracted from \cite{chiccoli1997phase}.}
    \label{Table1: critical temp and canted angles}
\end{table}

where $\lambda$ is the relative strength of $P_4$ with respect to $P_2$. It is a measure of the anisotropy in the system and can be positive or negative. The Legendre polynomial 
$P_4(z)=\left(35z^4-30z^2+3\right)/8$. With increasing strength of the $P_4$ term, the critical temperature $T_c^*$ increases and sharpens the first-order isotropic to nematic phase transition. Standard energy minimization reveals two classes of ground states for this model:
(i) For $\lambda < -0.3$, the global minimum of the energy results for $\theta_{ij}=\theta_c=\cos^{-1}\left[\left(15\lambda-6\right)/\left(35\lambda\right)\right]^{1/2}$ (or $\pi-\theta_c$). Thus, in the ground state, the nn nematic rods are aligned at an angle $\theta_{c}$, which is unique to each value of $\lambda$. 
We refer to these as {\it canted} states. (ii) For $\lambda\geq-0.3$, the global minimum of the energy is observed at $\theta_{ij} = \theta_{u}=0$ (or $\pi$), which produces the uniform {\it nematic} state. 
The variation of $\theta_c$ and $T_c^*$ for different values of $\lambda$ is specified in Table.~\ref{Table1: critical temp and canted angles}.

The canted phase has been observed experimentally in nematic LCs. The magnitudes of the second and fourth order Legendre order parameters $(\langle P_2\rangle$ and $\langle P_4\rangle)$ in LC samples can be obtained using Raman confocal microspectrometry, but these experimental measurements are rather difficult and scarce. Nevertheless, they have been reported in uniaxial LC p(DR1M-co-MMA), an azobenzene copolymer \cite{lagugne1998molecular, labarthet2000orientation}. The authors investigated the influence of $\langle P_4\rangle$ relative to $\langle P_2\rangle$, and observed that the angular distribution function $[P(\theta_{ij})$ vs. $\theta_{ij}]$ exhibits a peak at angles other than $\theta=0^\circ$ (or $180^\circ$).  However, the possible value of the model parameter $\lambda$ is not clear from these measurements. Thus, the GLL model provides a more versatile framework capable of describing a wider range of liquid-crystal phases, incorporating higher-order orientational interactions, and enabling systematic exploration of richer critical phenomena beyond the original LL model.
\subsection{Over-damped Langevin Dynamics} 
\subsubsection{Active Brownian Particle}
Typically, the ABP executes both a translational and a rotational motion. Assuming $\mathbf{r}_a(t)$ and $\hat{\mathbf{p}}_a(t)$ to be the instantaneous position and orientation of the ABP, the corresponding Langevin equations for translation and rotation are given by \cite{Risken1996}: 
\begin{eqnarray}
    \label{pos_abp_total}
    m\frac{d^2 \mathbf{r}_a(t)}{dt^2} &=& -\zeta_a\frac{d \mathbf{r}_a(t)}{dt}+ \nonumber \\
    &&F_0 \hat{\mathbf{p}}_a + \boldsymbol{\eta}(t),\\
    \label{ori_abp_total}
    I\left[\frac{d^2 \hat{\mathbf{p}}_a(t)}{dt^2} + \hat{\mathbf{p}}_a\left(\frac{d \hat{\mathbf{p}}_a(t)}{dt}\right)^2 \right] &=&-\xi_a\frac{d \hat{\mathbf{p}}_a(t)}{dt} + \nonumber \\ &&(\boldsymbol{\Omega}_{a}  + \boldsymbol{\Omega}_{an}) \times \hat{\boldsymbol{p}}_a.
\end{eqnarray}
Here, $m$ and $I$, are the mass and moment of inertia of the ABP, $\zeta_a$ and $\xi_a$ are the translational and rotational friction coefficients, $F_0$ is the active force on the ABP acting along its orientation, $\boldsymbol{\Omega_a}$ is the torque on the ABP due to its surrounding environments, and $\boldsymbol{\eta}$ and $\boldsymbol{\Omega_{an}}$ are the thermal noise in translation and rotation. Both thermal noises are Gaussian white noises with zero mean $\langle\eta_{\alpha}(t) \rangle=\langle \Omega_{a n}^{ \alpha}(t) \rangle=0$, following the correlations: $
    \langle \eta_{\alpha}(t) \eta_{\beta}(t') \rangle = 2 k_B T \zeta_a \delta_{\alpha \beta}\delta(t - t'),$ and $\langle \Omega_{a n }^{\alpha}(t) \Omega_{a n}^{ \beta}(t') \rangle = 2 k_B T \xi_a \delta_{\alpha \beta}  \delta(t - t') $, respectively.

Due to the negligible mass of most active particles, such as bacteria, the inertial terms in Eqs.~(\ref{pos_abp_total}) and (\ref{ori_abp_total}) can be neglected. In addition, they swim using their flagellar motion.  For such a scenario, the translational noise is insignificant compared to the rotational noise. This simplifies the dynamics to an over-damped motion, reducing Eqs.~(\ref{pos_abp_total}) and (\ref{ori_abp_total}) to \cite{toner2016following_ref02}: 
\begin{eqnarray}
    \label{pos_abp_od}
    \frac{d \mathbf{r}_a(t)}{dt} &=& \zeta_a^{-1} F_0 \hat{\mathbf{p}}_a\\
    \label{ori_abp_od}
    \frac{d \hat{\mathbf{p}}_a(t)}{dt} &=& \xi_a^{-1}(\boldsymbol{\Omega}_{a}  + \boldsymbol{\Omega}_{an})\times \hat{\boldsymbol{p}}_a
\end{eqnarray}

\subsubsection{Nematic Liquid Crystals}
For the LC background, we assume a system consisting of classical spins $\{\mathbf{r}_i\}$ of unit length on a simple cubic lattice. They are translationally immobile and have only rotational degrees of freedom. The dynamics of such spins can also be prescribed by the over-damped Langevin equation, analogous to the orientational dynamics of the ABP, which is given by the following \cite{toner2016following_ref02}: 
\begin{equation}
    \frac{d \hat{\mathbf{n}}_i(t)}{dt} = \xi^{-1} (\boldsymbol{\Omega}_{i} + \boldsymbol{\Omega}_{in}  ) \times \hat{\boldsymbol{n}}_i. 
    \label{ori_nem_od}
\end{equation}
Here, $\xi$ is the rotational friction coefficient of the nematogens and $\boldsymbol{\Omega}_{in}$ is the rotational noise with Gaussian white noise characteristics $\langle {\Omega}_{i n}^{\alpha} \rangle =0$ and $\langle {\Omega}_{i n}^{\alpha}(t) {\Omega}_{i n}^{\beta}(t') \rangle = 2 k_B T \xi \delta _{\alpha \beta} \delta(t - t')$ to ensure that the system is at temperature $T$ in the absence of the ABP. In addition, the torque on the $i^{th}$ nematogen by the nn is given by:
\begin{equation}
\label{nn LC interaction torque}
    \boldsymbol{\Omega}_i =  \epsilon \sum_{j} \left(3 + \frac{\lambda}{2} \left\{35(\hat{\mathbf{n}}_i \cdot \hat{\mathbf{n}}_{j}) -15 \right\} \right) (\hat{\mathbf{n}}_i \cdot \hat{\mathbf{n}}_{j})(\hat{\mathbf{n}}_i \times \hat{\mathbf{n}}_{j}).
\end{equation}
(Note that $\boldsymbol{\Omega}_i$ satisfies the force equation $\boldsymbol{\Omega}_i \times \hat{\mathbf{n}}_i = \mu_i\hat{\mathbf{n}}_i -  \partial H/\partial\hat{\mathbf{n}}_i$ where $\mu_i$ is the Lagrange multiplier that enforces the unit length restriction $|\hat{\mathbf{n}}_i |=1$.)

\subsubsection{ABP-NLC coupled system} 
In describing the dynamics of the ABP as it moves in the LC medium, we follow the assumptions made by Toner et al. The particle does not have long-term memory or an internal compass as it moves through an ordered nematic medium. As a consequence, the preference for any directional motion must arise from the local nematic director
$\hat{\mathbf{n}}_i(t)$ at the current location $\boldsymbol{{r}_a}(t)$ of the ABP. The set of over-damped Langevin equations for the combined system is as follows \cite{toner2016following_ref02}:
\begin{eqnarray}
    \label{Pos_ODE1}
    \frac{d \mathbf{r}_a(t)}{dt}&=& \zeta_a^{-1} F_0 \hat{\mathbf{p}}_a, \\
    \label{Ori_ABP_ODE2}
    \frac{d \hat{\mathbf{p}}_a(t)}{dt} &=& \xi_a^{-1}\left(\boldsymbol{\Omega}_{a}  + \boldsymbol{\Omega}_{an}\right)\times \hat{\boldsymbol{p}}_a,\\
    \label{Ori_LC_ODE3}
    \frac{d \hat{\mathbf{n}}_i(t)}{dt} &=& \xi^{-1} \left(\boldsymbol{\Omega}_{i} + \boldsymbol{\Omega}_{i,a} + \boldsymbol{\Omega}_{in}  \right) \times \hat{\boldsymbol{n}}_i. 
\end{eqnarray}
In Eq.~(\ref{Ori_LC_ODE3}), the resulting torque on the ABP due to the LC medium is given by $\boldsymbol{\Omega}_a = - \sum_i \boldsymbol{\Omega}_{i,a}$, where $\boldsymbol{\Omega}_{i, a}$ is the torque exerted by the ABP on the $i^{th}$ nematogen. Analogously to Eq.~(\ref{nn LC interaction torque}), it can be defined as:
\begin{eqnarray}
    \boldsymbol{\Omega}_{i,a} &=&  \epsilon_a  \left(3 + \frac{\lambda}{2} \{35(\hat{\mathbf{n}}_i \cdot\hat{\mathbf{p}}_a  ) -15 \} \right)  (\hat{\mathbf{n}}_i \cdot\hat{\mathbf{p}}_a)\nonumber \\ &&(\hat{\mathbf{n}}_i \times\hat{\mathbf{p}}_a) g(\mathbf{r}_i - \mathbf{r}_a), 
    \label{ABP_LC_Int_Torq}
\end{eqnarray}
where $\epsilon_a$ is the strength of the interaction between ABP and nematogen, and $g(r) = \exp[-(r/\sigma)^2]$ is the additional term that incorporates spatial dependency, with a characteristic decay length $\sigma$. In general, bacteria use flagella for motility and $\sigma$ defines the average length of these flagella, typically longer than the lattice constant to overcome the discretization of the medium. Typically, ABP aligns with the local director, which leads us to set $\lambda=0$ in the ABP-LC interactions. 

Taking into account the lattice constant $a$ and the rotational diffusion constant $D_r = k_BT/\xi$ for LC molecules, we can introduce several dimensionless parameters that simplify the over-damped Langevin equations. To do this, we replace $t=\tau/D_r$, $F_0=k_BT f_0/a$, and $\boldsymbol{\Omega}_x = \boldsymbol{\omega}_x k_BT$, where $\tau$ is the reduced time, $f_0$ is the Peclet number and $\boldsymbol{\omega}_x$ is the reduced torque on the ABP or nematogen, as a case might be. We also define two ratios of friction coefficients $ \tilde{\zeta}_a  = a^2 \zeta_a / \xi $ and $\tilde{\xi}_a = \xi_a/ \xi$. Furthermore, assuming $l$ to be the size ratio of the ABP and LC molecules and using Stokes law for spherical particles, it is easy to express $\tilde{\zeta}_a = 3l/4$ and $\tilde \xi_a = l^3$. Using these dimensionless parameters, the over-damped equations can now be written as:
\begin{eqnarray}
    \label{scaled_odle1}
     \frac{d(\mathbf{r}_a/a)}{d \tau} &=& \tilde{\zeta}^{-1}_a f_0 \hat{\mathbf{p}}_a, \\
    \label{scaled_odle2}
    \frac{d \hat{\mathbf{p}}_a }{d \tau}  &=& \tilde{\xi}^{-1}_a \left( \boldsymbol{\omega}_a +  \boldsymbol{\omega}_{an} \right) \times \hat{\mathbf{p}}_a,\\
    \label{scaled_odle3}
    \frac{d \hat{\mathbf{n}}_i }{d \tau}  &=& \left( \boldsymbol{\omega}_i +  \boldsymbol{\omega}_{i,a} +  \boldsymbol{\omega}_{in} \right) \times  \hat{\mathbf{n}}_i, 
\end{eqnarray}
with the scaled Gaussian noise in the two components of the coupled system obeying $\langle {\omega}_{an}^{\alpha} \rangle =0$, $\langle {\omega}_{a n}^{\alpha}(\tau_1) {\omega}_{a n}^{\beta}(\tau_2) \rangle = 2 \tilde{\xi}_a \delta_{\alpha \beta}\delta(\tau_1 - \tau_2)$ and 
$\langle {\omega}_{i n}^{\alpha} \rangle =0$,
$\langle {\omega}_{i n}^{\alpha}(\tau_1) {\omega}_{i n}^{\beta}(\tau_2) \rangle = 2 \delta _{\alpha \beta} \delta(\tau_1 - \tau_2)$. The set of Eqs.~(\ref{scaled_odle1})-(\ref{scaled_odle3}) provide the framework for investigating the consequences of the LC textures on the trajectories of the ABP.

\section{Numerical Results}
\label{Results}
\subsection{Computational Details}  
The nematogens occupy sites in a cubic lattice $L^3$ with $L=64$ that follows periodic boundary conditions.
Our model incorporates two types of interactions: (i) nn interactions among LC molecules and (ii) interactions between the ABP and the surrounding LC medium, which decay exponentially with the separation distance. Consequently, we restrict the ABP-LC interaction to nematogens within a volume $V_I$, defined as a sphere of radius $7a$. We verified that further increasing this cutoff radius does not affect our numerical results. The evolution of the coupled ABP-LC system is studied using a simple Euler integration scheme to solve Eqs.~(\ref{scaled_odle1})- (\ref{scaled_odle3}) with $d\tau = 10^{-3}$.
The values of the model parameters are as used in the study by Toner et al. where $f_0=5 $, $l=5$ and $ \sigma =2a$. The latter corresponds to the distance where the interaction decays to $1/e$. For simplicity, we set $a=1$ and $\epsilon=\epsilon_a =1$. The standard deviations of Gaussian white noise are $\sqrt{2\tilde{\xi}_a^{-1}}$ for ABP and $\sqrt{2}$ for LC molecules. Initially, all nematogens are aligned along the $z$-axis. With time, the director $\hat{\bf n}=\sum_{i=1}^N \hat{\bf n}_i/N$ fluctuates around $z$ due to the noise and motion of the ABP. The coupled system is allowed to evolve for $\tau =100$ steps with the ABP fixed in the center of the lattice. The particle is then released, and the system is equilibrated for  $\tau =1000$  before data collection over an interval of $\tau =9000$.
The resultant torque $\boldsymbol{\Omega}_a$ on the ABP due to the LC medium [provided by Eq.~(\ref{Ori_LC_ODE3})] is also obtained considering nematogens within the spherical volume $V_I$. 
Initially, the simulation was implemented using a serial C++ code in which all interactions were evaluated sequentially. This implementation required approximately $15{,}810$ seconds to complete the $10^5$ time steps. Subsequently, we developed a parallel implementation based on CUDA in which nn LC-LC interactions are computed in parallel for all LC molecules, and ABP-LC interactions are also fully parallelized. The CUDA implementation reduces the runtime to approximately $50$ seconds for $10^5$ time steps, yielding a speedup of nearly a factor of $300$ compared to the serial C++ code.
\begin{figure}[H]
    \centering
    \includegraphics[width=1\linewidth]{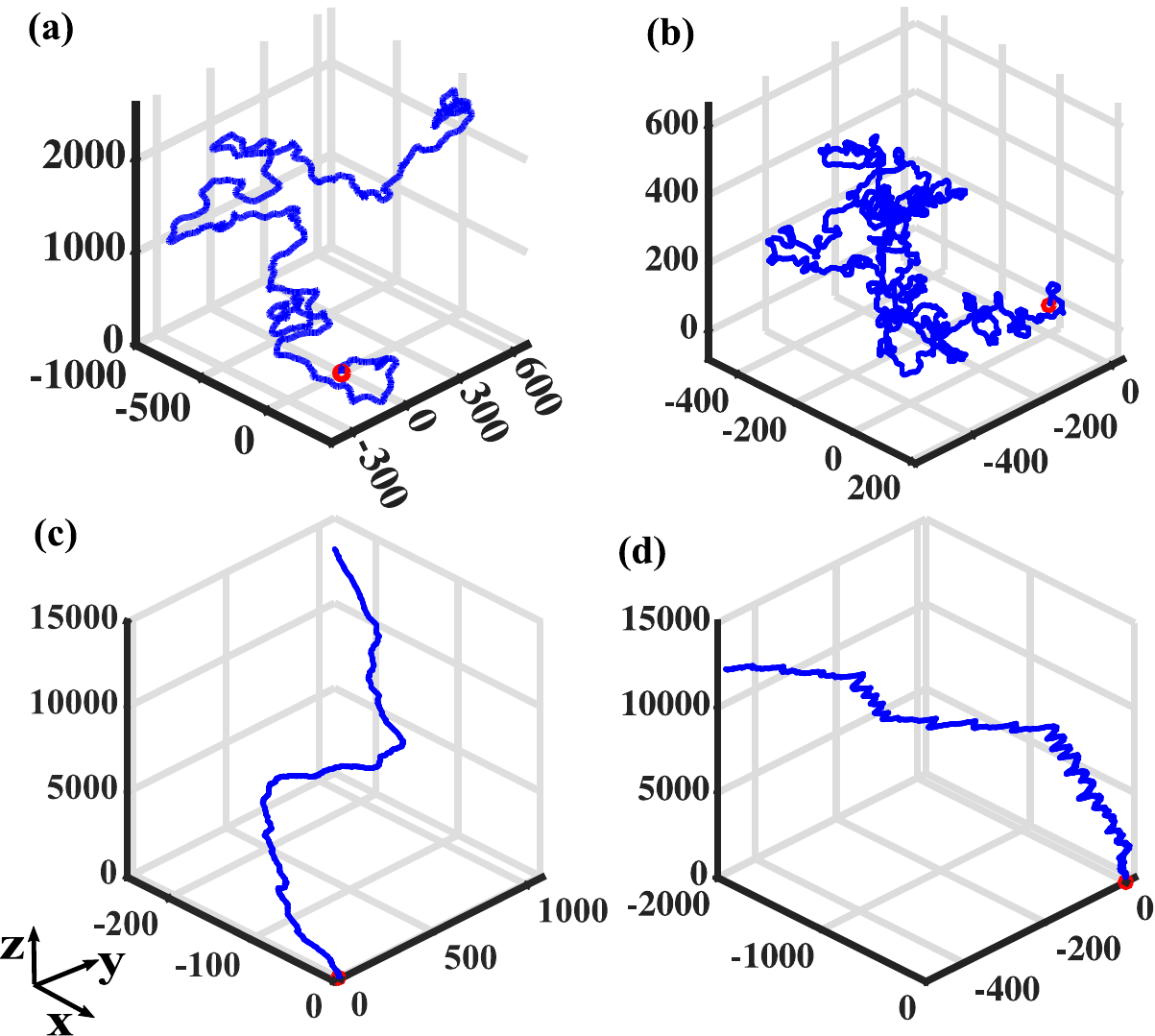}
    
    \caption{Trajectories traced by the ABP in (a) free space with no medium, (b) isotropic phase $(\lambda=0, T^*> T^*_c)$, (c) nematic phase $(\lambda=0, T^*=T^*_c/2)$, and (d) canted phase $(\lambda=-3, T^*=T^*_c/2)$. The red circle indicates the starting point. The director $\hat{\mathbf{n}}$ of the ordered LC medium is oriented nearly parallel to the z-axis, exhibiting only very small deviations. }
    \label{fig:trajectories}
\end{figure}
\begin{figure*}[ht]
    \centering
    \includegraphics[width=1\linewidth]{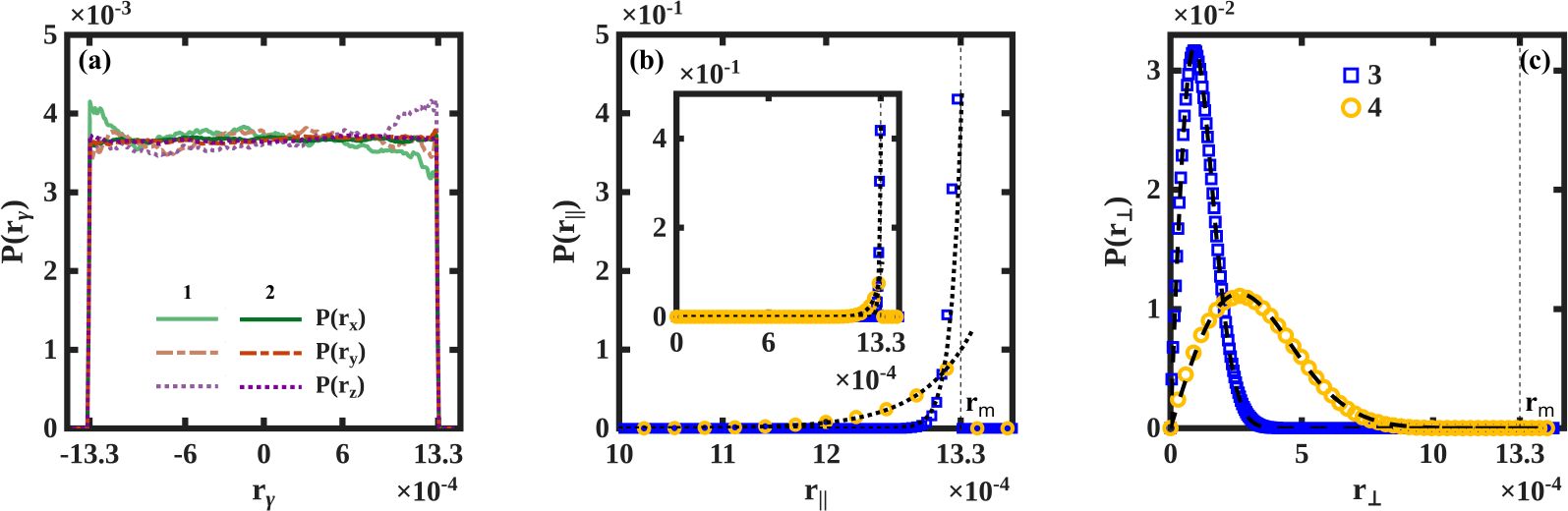}
    \captionsetup{justification=raggedright,
    singlelinecheck=false}
    \caption{SSD for the different cases: (a) $x$, $y$, and $z$ components for the walks in {\it Case 1} and {\it Case 2}. (b) Parallel component ($r_{||}$) for {\it Cases 3} and {\it Case 4}. The dashed lines are fits to $f(r_{||}) \sim \exp(ar_{||})$ with $a = 1.35 \times10^5$ ({\it Case 3}) and $a = 2\times10^4$ ({\it Case 4}). (d) Perpendicular component ($r_{\perp}$) for {\it Case 3} and {\it Case 4}. The dashed lines are fits to the Rayleigh distribution provided in Eq.~(18) with $\sigma = 9.23\times10^{-5}$ ({\it Case 3}) and $\sigma = 2.61\times10^{-4}$ ({\it Case 4}).}
    \label{fig:step-size}
\end{figure*}
We investigate the following cases, choosing prototypical values of $\lambda$ and $T$:
(1) ABP in free space. This is tantamount to solving Eqs.~(\ref{scaled_odle1}) and (\ref{scaled_odle2}) setting the torque $\boldsymbol{w}_{a}=0$, so that no external rotational influences are introduced. The (unscaled) noise correlations are chosen to be proportional to $T=0.56$.
(2) ABP in an isotropic LC medium for $\lambda=0$ and temperature $T=5$; 
(3) ABP in a nematic LC medium with $\lambda=0$ and $T=T_c(\lambda=0)/2=0.56$;
(4) ABP in a canted medium with for $\lambda=-3.0$ and $T=T_c(\lambda=-3)/2=0.25$. The motion of the particle is assumed to be along the local nematic director. Consequently $\lambda = 0$ in Eq.~(\ref{ABP_LC_Int_Torq}). 

To begin with, it is useful to know the textures of the LC medium for different choices of $\lambda$ and $T$. Fig.~1(a) shows the $xz$ slice of the isotropic phase for $\lambda = 0$ and $T=5$.  Fig.~1(b) shows the corresponding nematic morphology obtained for $\lambda = 0$ and $T = 0.56$. The canted morphology at $y=0$ for $T=0.25$ is shown in Fig.~1(c). This choice of $\lambda$ yields a canting angle $\theta_c=45.82^\circ$, see Table~I. (There are fluctuations about this value due to thermal randomization.) The values of the ensemble averaged orientational order parameter $\color{teal}{\mathcal{S}}$, as defined in Eq.~(\ref{equation_2}), are also specified for each of the textures.
Few important observations regarding the canted state are in order. Because the intermolecular potential is minimized at $\theta _{c}$, any pair of neighboring rods can tilt away from each other along a three-dimensional cone of degeneracy. This geometric freedom yields structurally complex, local ``zigzag'' or helical configurations (illustrated by the cyan and pink shading in the inset). Consequently, the canted morphology fractures into a polycrystalline maze of distinct structural domains separated by sharp, well-defined ``interfacial defects'' \cite{Birdi_2020_ref03}. They stands in stark contrast to standard nematic morphologies, which are instead characterized by continuous, extended disclination lines or ``string defects''  \cite{vats2021, Birdi_2020_ref03}. We will demonstrate that this fundamental difference in defect topology - sharp interfacial domain walls versus continuous line defects - profoundly alters the transport characteristics of the ABP.

Let us now study the motion of the ABP through the LC medium by numerically solving Eqs.~(\ref{scaled_odle1})-(\ref{scaled_odle3}). The evaluations are expected to reflect the interplay of active propulsion, thermal fluctuations, and the directional order imposed by the LC medium.  Fig.~(\ref{fig:trajectories}) shows the  trajectory traced for (a) {\it Case 1}, (b) {\it Case 2}, (c) {\it Case 3}, and (d) {\it Case 4}. 
In free space, the particle exhibits the characteristic Brownian trail. In the isotropic medium, the LC molecules are randomly oriented. Following the local director, the ABP is trapped in localized regions before exiting them, as observed in Fig.~\ref{fig:trajectories}(b). Figure \ref{fig:trajectories}(c) shows that in the nematic medium, the LC molecules guide the ABP into a directed motion. In the canted medium, the background LC is quite complex, as seen in Fig.~1(c). Due to the nature of the coupling term, the ABP aligns with the local director. As observed in Fig.~\ref{fig:trajectories}(d), the motion is directed, but much slower than in the nematic background. Thus, there are subtle differences in the trajectories, which we identify in forthcoming evaluations.

To understand the diffusion of ABP in the different textures, we examine its fundamental microscopic elements, the individual steps $\vec{\bf r}$ that are taken as time evolves. If we follow Eq.~(\ref{scaled_odle1}), the magnitude of step $|dr| = r_m = f_0d\tau/\zeta_a=1.33\times 10^{-3}$. So, the SSD is a delta-function: $P(r)=\delta(r_m-r)$. However, the parallel and perpendicular components vary as a result of the anisotropy of the medium.  Fig.~\ref{fig:step-size}(a) that shows $P(r_{\gamma})$. vs. $r_{\gamma}$ ($\gamma=x,y,z$) for {\it Case 1} and {\it Case 2}. As expected, the distributions are uniform since the ABP executes a Brownian walk in free space and in the isotropic phase of the LC. 
For {\it Case 3} and {\it Case 4}, the active particle moves in anisotropic textures. In this context, it is relevant to consider the parallel and perpendicular components, $r_{||}$ and $r_{\perp}$, with respect to $\hat{n}$.  
Fig. ~\ref{fig:step-size}(b) shows $P(r_{||})$ vs. $r_{||}$ for the nematic and canted phases. The dashed curves represent fits to a left-tailed exponential distribution [$P(r_{||}) \sim \exp(-a r_{||})$]. It reflects persistence along $\hat{\bf n}$, and the steps $r<r_m$ are due to random noise fluctuations. The latter have a larger influence on the canted medium, possibly due to the lower orientational order. The distributions for $P(r_{\perp})$ vs. $r_{\perp}$ are shown in Fig.~\ref{fig:step-size}(c). The dashed lines are the fits to the Rayleigh distribution given by:
\begin{equation}
    P(r_{\perp}) = \frac{r_{\perp}}{ \sigma^2}\exp{\left( -\frac{r_{\perp}^2}{2\sigma^2} \right)}, 
    \label{Rayleigh PDF}
\end{equation}
with mean $\sigma\sqrt{\pi/2}$ and variance $(4-\pi)\sigma^2/2$. Note that the Rayleigh distribution models the magnitude of a two-dimensional vector ($r_{\perp} = \sqrt{x^2+y^2}$) whose components are independent and identically distributed random variables drawn from a Gaussian distribution with mean 0 and variance $\sigma$. Although the Rayleigh distribution is expected for $r_{\perp}$ in the nematic phase, it is interesting to note that even in {\it Case 4}, the steps are consistent along $\hat{\bf n}$ and are isotropic in the $xy$ plane. The corresponding distributions, though of the same forms, have smaller magnitudes, are wider, and have longer tails reflecting the higher symmetry of the canted medium. 

Observing the Rayleigh distribution in the $xy$-plane alongside an exponential distribution in the $z$-direction typically indicates an anisotropic system, one in which vertical and horizontal movements are governed by different physical constraints. The horizontal $xy$ motion remains free and diffusive, characterized by a Brownian motion that explores an open plane, leading to a Rayleigh distribution of the radial distance. The vertical motion, on the other hand, is pinned and biased leading to an exponential profile in the $z$ coordinate, often seen in the presence of strong external fields such as gravity or electric, or highly directional molecular interactions. The distributions confirm that the dynamics in the $xy$ and $z$ directions are statistically independent and are governed by separate physical potentials or boundary conditions. In the context of our discussion related to the emergence of Goldstone modes, the exponential $z$ profile implies that a strong energy "gap" or bias has been imposed, suppressing free fluctuations.

A quantitative measure of the roughness of the trajectories can be obtained from the Hurst exponent, which is determined by evaluating the Brownian trace. The instantaneous displacement of the ABP at time $t$ is expressed as $\left|\Delta \boldsymbol{\Vec{r}}(\tau) \right| =\left| \boldsymbol{\Vec{r}}(\tau) - \boldsymbol{\Vec{r}}(0) \right|$. Taking into account pairs $\left(\tau_i, \tau_j\right)$ such that $\left|\tau_i-\tau_j\right|=\tau_w$, the Brownian trace is defined by \cite{paul1997addison}:
\begin{equation}
    {\Delta B(\tau_w)= \langle \overline{\left|\Delta \boldsymbol{\Vec{r}}(\tau+ \tau_w)\right| - \left|\Delta \boldsymbol{\Vec{r}} (\tau) \right|}\rangle}, 
    \label{Brownian_trace_eqn}
\end{equation}
where the overline indicates an average over all pairs $\left(\tau_i, \tau_j\right)$ and the angular brackets denote an ensemble average. Generally, $\Delta B(\tau_w) \propto \tau_w^H$  with $H$ as the Hurst exponent. On a related note, persistence refers to the particle continuing in the same direction, whereas anti-persistence indicates a reversal. A value of $H<0.5$ signifies an anti-persistent sub-diffusive path,  $H=0.5$ indicates Brownian motion, and $H > 0.5$ suggests a persistent, super-diffusive ballistic walk. A value of $H=1$ signifies a straight and smooth ballistic walk.
The Hurst exponent can also be utilized to determine the fractal dimension of the trajectory $d_f = d - H$ \cite{paul1997addison}. The fractal dimension quantifies how space-filling the path is. The lower values of $d_f$ correspond to a smoother, more correlated and less space-filling trajectory. Higher values of $d_f$ on the other hand indicate a rougher, more random, and space-filling trajectory. On a related note, persistence refers to the particle continuing in the same direction, whereas anti-persistence indicates a reversal. A value of $H<0.5$ signifies an anti-persistent sub-diffusive path,  $H=0.5$ indicates Brownian motion, and $H>0.5$ suggests a persistent, super-diffusive ballistic walk. A value of $H=1$ signifies a straight and smooth ballistic walk.
\begin{figure*}[hbt]
    \centering
    \includegraphics[width=1\linewidth]{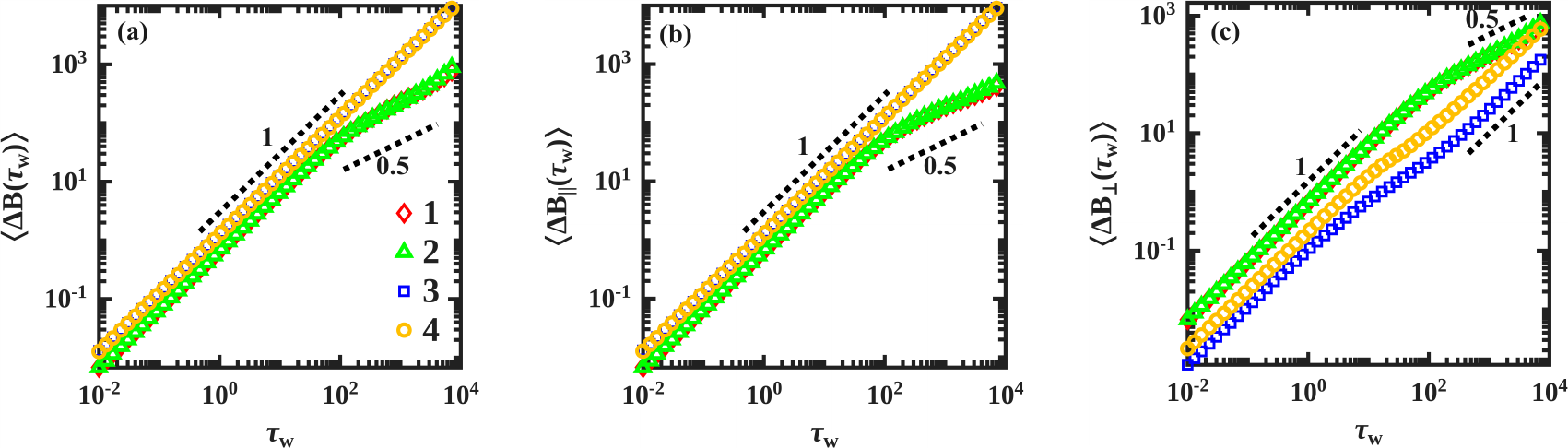}
    \captionsetup{justification=raggedright,
    singlelinecheck=false}
    \caption{(a) Evaluation of the Hurst exponent (a) using Eq.~(19), (b) from parallel components $r_{\parallel}$, (c) from perpendicular components $r_{\perp}$. The dashed lines with specified slopes are a guide to the eye.}
    \label{fig:Brownian_trace}
\end{figure*}
The evaluation of the Brownian trace and the corresponding Hurst exponent for all four cases is shown in Fig.~\ref{fig:Brownian_trace}. The dashed lines with specified slopes are a guide for the eye. As seen in the sub-figure (a), the ABP trajectories in free space and in the isotropic LC medium indicate persistence and smoothness for a short time scale but become ragged and uncorrelated at larger times with $d_f=1.5$, which is typical of Brownian walks. The motion in the nematic and canted phases is ballistic at all observation times. To understand the consequences of the anisotropy of the LC medium on the motion of the particle, we also evaluated the Hurst exponent for the parallel and perpendicular components of the steps with respect to $\hat{\mathbf{n}}$ in Figs.~\ref{fig:Brownian_trace}(b) and \ref{fig:Brownian_trace}(c). Surprisingly, the trajectories are not fractal in the canted phase inspite of it's complex texture. Furthermore, the Brownian trace derived from the transverse component in Fig. \ref{fig:Brownian_trace}(c) displays a distinct bilinear scaling profile separated by a ``kink''. This feature originates from the competition between the ABP’s self-propulsion and the restorative torque of the nematic medium \cite{bechinger2016active}. When the particle’s activity drives it perpendicular to the director, it must overcome a significant energy barrier to reorient against the nematic alignment. The initial linear regime ($H\approx 1$) corresponds to localized propulsion within a cage defined by the local director field. The crossover represents the activation point where the particle escapes this confinement, either through sustained self-propulsion or thermal fluctuations, leading to a second linear regime that characterizes inter-cage hopping or layer-to-layer transport.
\begin{figure}[H]
    \centering
    \includegraphics[width=0.5\linewidth]{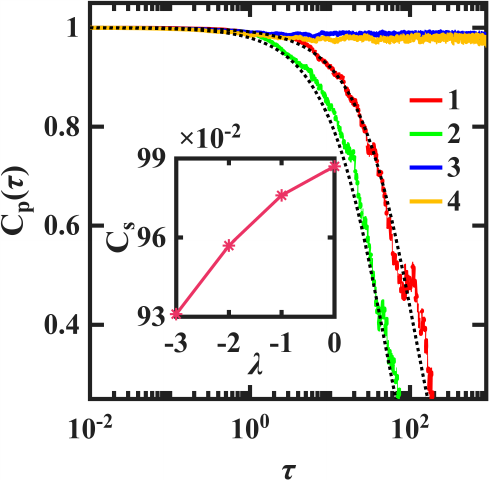}
    \caption{The orientation auto-correlation function of the ABP for the different cases under consideration. The dashed lines are fits to the exponential form $f(\tau) \sim \exp(-b\tau)$ with $b = 8.32\times10^{-3}$ ({\it Case 1}) and $b = 7.53\times10^{-2}$ ({\it Case 2}). For the ordered nematic phases, the correlation saturates to a constant value $C_s$. The inset shows the variation of $C_s$ with $\lambda$.}
    \label{fig:memory}
\end{figure}
We also investigated whether the orientation of ABP is well maintained in the LC medium by evaluating the orientation autocorrelation function $C_{p}(\tau)=\langle \hat{\mathbf {p}}(0)\cdot \hat{\mathbf {p}}(\tau)\rangle$. Fig.~\ref{fig:memory} shows $C_p(\tau)$ vs. $\tau$ for the different cases. The ABP rapidly loses memory in {\it Cases 1} and {\it 2}. The dashed lines are exponential fits $P(\tau)\sim \exp(-b \tau)$ with $b \simeq 0.008$ and $b \simeq 0.075$, respectively. The orientation is well maintained in the nematic medium. In the canted medium, $C_p(\tau)$ saturates to a constant value $C_{s}$ that depends on $\theta_c(\lambda)$. The latter, as discussed in Sec.~IIA, depends on the relative strength of the $P_4\left(\cos\theta\right)$ term in the GLL Hamiltonian [Eq.~(\ref{equation_3})] that takes into account the higher order orientational correlations beyond those captured by $P_2\left(\cos\theta\right)$. The inset shows the variation of the saturation value $C_s(\lambda)$ vs. $\lambda$. The slight decoherence in the nematic phase is attributed to the finite values of temperature (noise). Although the ABP meanders in the canted phase as a result of the coupling with the LC molecules, surprisingly the autocorrelation is well-sustained even in this complex phase.  

Finally, we evaluated the MSD of the ABP, which is a quantitative measure of how far, on average, a particle has moved from its initial position over time and is defined by the following:
\begin{equation}
       \langle \Delta R^{2}(\tau)\rangle = \langle \mid \mathbf{r}(\tau) - \mathbf{r}(0)\mid^2\rangle, 
       \label{MSD}
\end{equation}
where $\langle\cdot\cdot\rangle$ denotes an ensemble average over particle trajectories.
The MSD, through its scaling over time $\langle \Delta R^{2}(\tau)\rangle \propto \tau^{\alpha}$, characterizes the diffusion process \cite{saxton1997single}. A linear dependence ($\alpha = 1$) corresponds to normal diffusion, typical of Brownian motion that is random and collision dominated.  Deviations ($\alpha \neq 1$) indicate an anomalous diffusion. Sub-diffusive motion ($\alpha < 1$) arises when particles are transiently trapped or subject to memory effects, while super-diffusion ($\alpha > 1$) can result from long-range correlations or L\'evy walk–like dynamics. In this case, the walk is ballistic, suggesting straight-line or inertia-dominated motion. 

\begin{figure}[H]
    \centering
    \includegraphics[width= 1\linewidth]{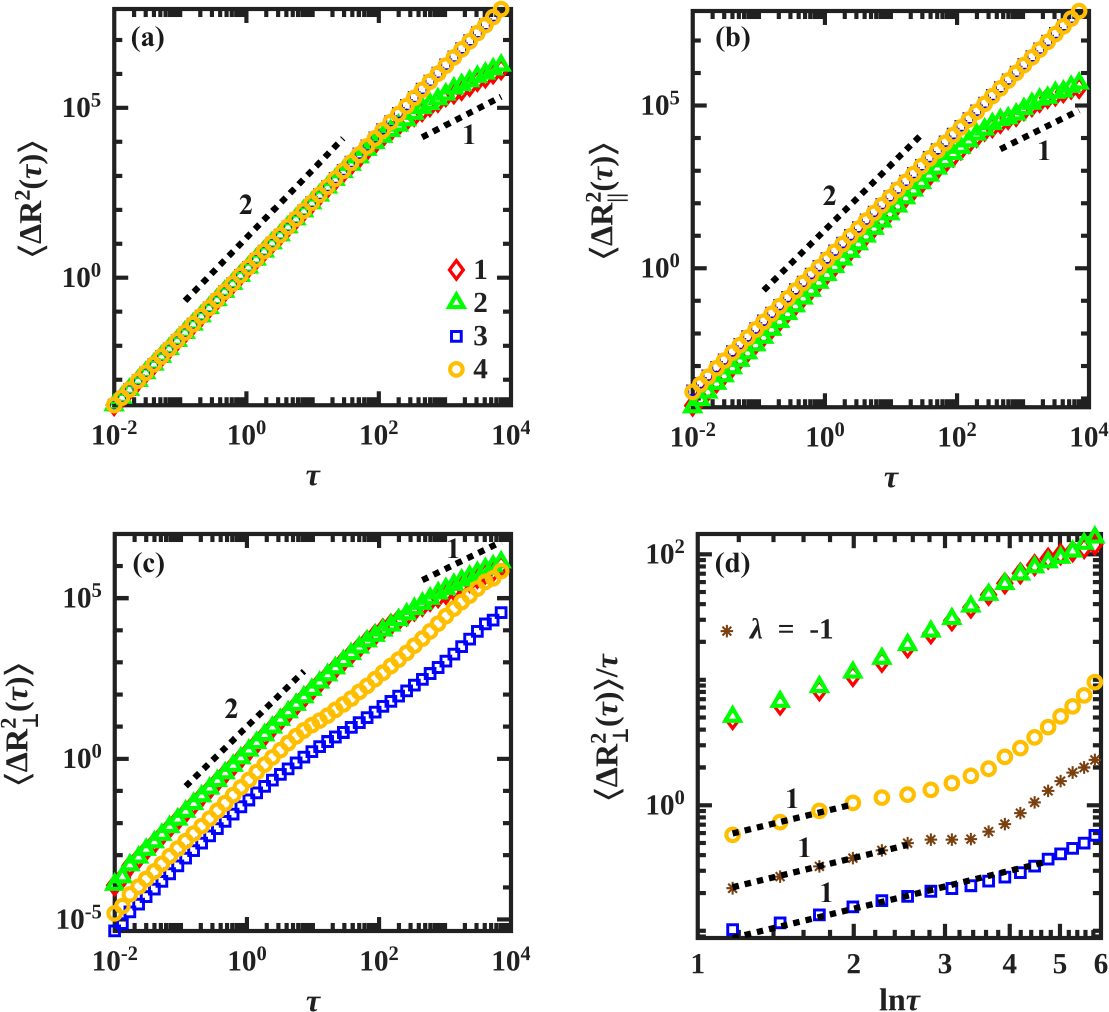}
    \caption{(a) Evolution of the MSD  for the different cases; (b) for the parallel displacement; (c) for perpendicular displacement. The $t\ln t$ corrections in the perpendicular component are investigated in (d). In each sub-figure, dashed lines with indicated slopes are a guide to the eye.
    }
    \label{fig:msd}
\end{figure}

Fig.~\ref{fig:msd}(a) shows the MSD $\langle \Delta R^{2}(\tau)\rangle$ vs. $\tau$ for all four cases that we have considered. We also show data for $\lambda=-1$ to systematically check the effect of the canted angle $\theta_c$ on the suggested logarithmic corrections. 
The corresponding plots evaluated from the components of the steps along $\hat{\bf n}$ [$\langle \Delta R^{2}_{||}(\tau)\rangle$ vs. $\tau$] and perpendicular to $\hat{\bf n}$ [$\langle \Delta R^{2}_{\perp}(\tau)\rangle$ vs. $\tau$] are shown in Figs.~\ref{fig:msd}(b) and (c). The dashed lines with the indicated slopes are a guide for the eye. In all cases in (a) and (b), the initial data fit to $\alpha = 2$, indicating ballistic motion. There is a cross-over to a diffusive regime ($\alpha = 1$) at late times for {\it Case 1} and {\it Case 2}. This behavior is typical of stochastic, fluctuating trajectories that are generated due to a combination of diffusion and self-propulsion of the ABP. In the nematic and canted medium,  the ABP  continues to exhibit persistent ballistic motion even at late times. 
The MSD evaluation of the transverse component shown in Fig.~\ref{fig:msd}(c). 
To verify logarithmic corrections, we plot in Fig.~\ref{fig:msd}(d), $\langle \Delta R^{2}(\tau)\rangle$ vs. $\ln \tau $ on a double-logarithmic scale. The relevant signature will be a linear fit of slope 1 to the data. Clearly, there are no logarithmic corrections for {\it Case 1} and {\it Case 2}. The dashed lines on the data sets corresponding to $\lambda=0$, -1, -3 are a guide to the eye. The corrections in the nematic medium ($\lambda = 0$) are prominent and consistent with the observation made by Toner et al. The behavior in the canted phases ($\lambda= -1$ and $-3$) is different.
The logarithmic regime progressively shrinks with increasingly negative values of $\lambda$, suggesting a systematic suppression of the long-wavelength Goldstone modes that typically fuel active super-diffusion.

To elucidate the physical mechanism driving the loss of these logarithmic corrections in the canted phase, we examine the underlying topological defect configurations. Fig.~7(a) 
isolates the interfacial boundaries that separate the adjacent canted domains for $\tau  = 100 $. If $\theta_{ij}$ deviates significantly from the ideal ground-state canting angle ($\theta_{ij} \neq \theta_{c} \pm 10^o$), the corresponding bond midpoint is highlighted in red. The canted morphology is visible to be fractured into a maze of distinct structural domains bounded by sharp and well-defined interfacial defects \cite{Birdi_2020_ref03}. Similarly, starting with an isotropic state, Fig.~7(b) depicts the string defects that emerge in the nematic phase ($T=0.56$) at $\tau = 20$. These were obtained by considering all possible square plaquettes in the cubic lattice and checking whether the nematic rods rotate by $180^o$ \cite{Birdi_2020_ref03}. While the string defects are annihilated as the system coarsens, the canted medium continues to exhibit a maze of distinct structural domains bounded by sharp interfacial defects even at late times.

\begin{figure}[H]
    \centering
    \includegraphics[width=1\linewidth]{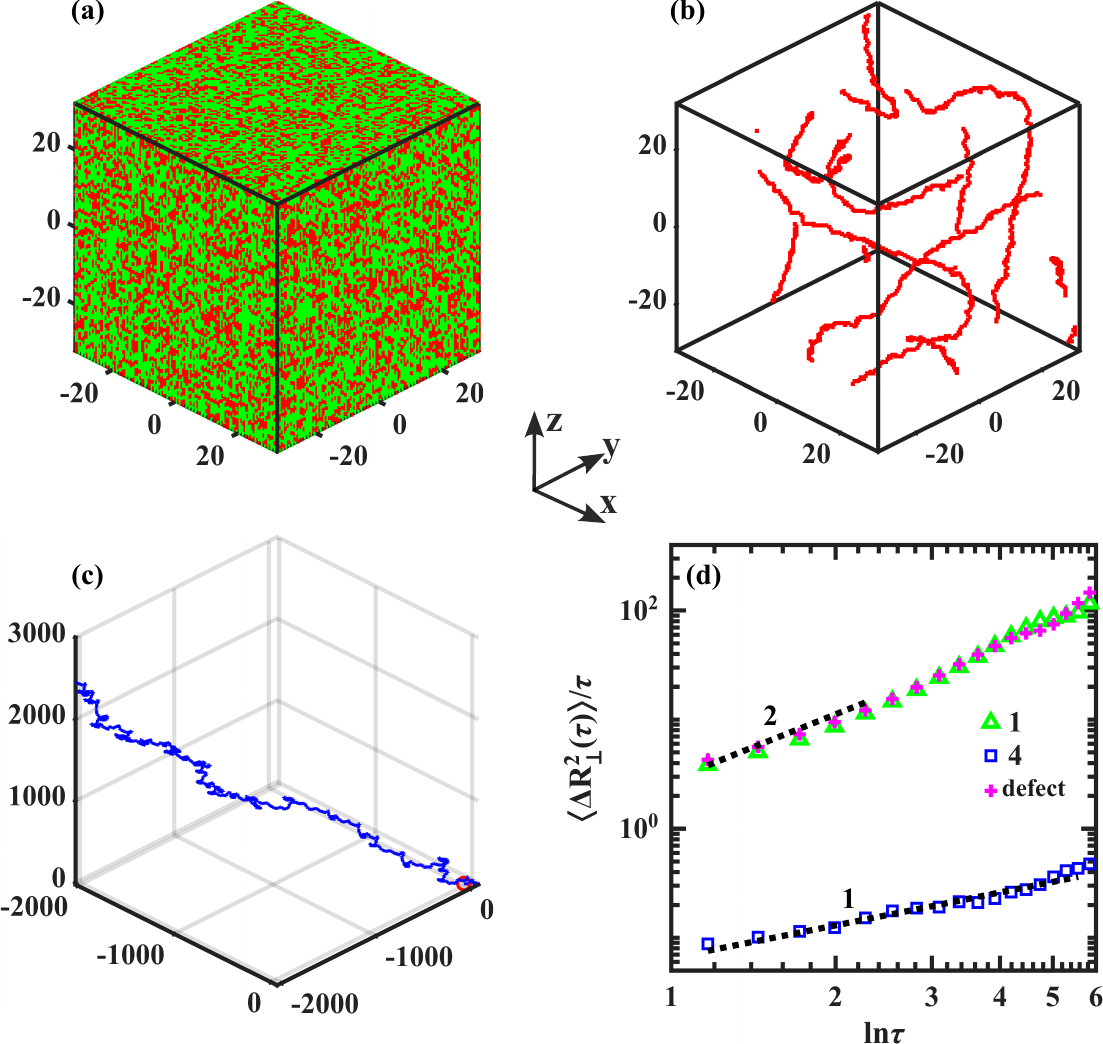}
    \caption{(a) Interfacial defects (red) in the canted phase.  (b) String defects in the nematic phase at $\tau = 20$ from an initial isotropic state. (c) Trajectory of the ABP in the nematic LC medium with string defects shown in (b). (d) Perpendicular component of the MSD averaged over several trajectories traced in the defect morphology of (b) denoted by {\color{magenta}+}.}
    \label{fig:defects}
\end{figure}

Note that in our analysis, the ABP moves in an aligned nematic, as was assumed in the early papers on this topic \cite{toner2016following_ref02, toner2018_smectic}. For completeness, we also checked for the influence of string defects on the ABP trajectory, as they are ubiquitous at finite temperatures. Fig.~\ref{fig:defects}(c) shows a prototypical trajectory traced by the ABP in the environment depicted in Fig.~\ref{fig:defects}(b). Although dominated by the orientation of the nematic director $\hat{\bf n}$, the trajectory is rough compared to that shown in Fig.~\ref{fig:trajectories}(c). Note that in our analysis, the ABP moves in an aligned nematic medium, as assumed in \cite{toner2016following_ref02, toner2018_smectic}.  Fig.~\ref{fig:defects}(d) shows $\langle \Delta R^{2}(\tau)\rangle$ vs. $\ln \tau $ (denoted by +) obtained from 20 trajectories traced in the morphology of Fig.~\ref{fig:defects}(b). Clearly, the $t\ln t$ corrections are absent, indicating that the super-diffusive kick to the ABP is lost in the presence of vortex strings.
A true hydrodynamic Goldstone mode requires a globally uniform and continuous symmetry-broken state where a long-wavelength fluctuation ($k \to 0$) costs zero energy. Although the uniform nematic state preserves this condition across macroscopic scales, the fragmented nature of the canted phase breaks the continuity of the director field. Because these canted domains are randomly oriented and finite in size, the continuous rotational symmetry is effectively broken on length scales that exceed the characteristic domain size ($\ell$). Consequently, the gapless Goldstone modes that propagate freely in the fully aligned nematic phase are structurally choked in the canted phase due to severe {\it Porod} scattering and pinning at the sharp domain boundaries \cite{Birdi_2020_ref03}. From these geometric arguments, it is evident that the compartmentalization of the canted phase introduces a finite mass gap to the orientational fluctuations, rendering the hydrodynamic modes short-lived. This structural confinement ultimately curtails the logarithmic super-diffusive regime, the extent of which remains governed by the specific domain morphology dictated by $\lambda$ (or $\theta_c$). Similarly, the presence of topological defects also breaks long-wavelength fluctuations.

\section{Summary and Discussion}
\label{Summary and Discussion}
We have investigated the dynamics of an ABP in a non-Newtonian LC medium modeled by the GLL model. In addition to the second order Legendre polynomial $P_2$ that captures the uniaxial orientational order, the GLL model also includes $P_4$ which incorporates higher order orientational correlations responsible for many of the experimentally observed biaxial or higher order nematic phases. The GLL model, by tuning the relative strengths of the $P_2$, $P_4$ interactions, leads to nematic and canted textures. The interaction between the ABP and the LC medium is chosen to support their co-alignment. Over-damped Langevin formalisms were employed to represent the ABP dynamics and the evolution of the LC orientations. To understand the motion of ABP in these environments, we evaluated quantifiers such as SSD, Brownian trace, fractal dimension, orientational autocorrelation, and MSD. Each of these measures offers complementary information. The fractal dimension characterizes spatial exploration and aggregation: higher $d_f$ corresponds to space-filling trajectories and compact aggregates, while lower $d_f$ indicates branched or sparse structures. The SSD reflects the effective ``rules of motion'' that govern the ABPs. Narrow distributions such as the Gaussian or exponential indicate smooth, localized trajectories dominated by small steps, consistent with Brownian diffusion. In contrast, heavy-tailed Levy distributions suggest erratic dynamics characterized by frequent short steps interspersed with rare but significant long flights, a signature of strong fluctuations in the driving forces. The MSD quantifies the temporal spread of the particle’s position distribution and serves as a key diagnostic of diffusion behavior and transport efficiency.

We summarize the result of our investigations below: 
(i) Isotropic Phase: Analysis of the Brownian trace indicates rough, space-filling trajectories with a high fractal dimension $(d_f = 3.5)$. The orientational autocorrelation decays exponentially to zero, confirming a rapid loss of directional memory. Because the medium is macroscopically isotropic, the SSD is completely uniform across all spatial directions. The MSD exhibits a short-time ballistic regime that smoothly transitions into standard Fickian diffusion at longer timescales, consistent with classical passive Brownian motion.
(ii) Nematic phase: The uniform, anisotropic alignment of the nematogens preferentially guides the ABP along the global director ${\hat{\bf n}}$ sustaining extended ballistic trajectories that remain smooth and non-fractal across all observed timescales. The particle maintains a long-lived memory of its direction of motion. The parallel component of SSD follows a left-tailed exponential distribution along ${\hat{\bf n}}$  while the perpendicular component obeys a Rayleigh distribution. Crucially, while the parallel MSD scales ballistically at all times, the perpendicular component reveals prominent logarithmic ($t \ln t$) corrections - a signature of anomalous active super-diffusion fuelled by the gapless, transverse Goldstone modes of the nematic matrix.
(iii) Canted phase: Despite the complex spatial arrangements of the nematogens, the ABP maintains smooth, non-fractal trajectories. Interestingly, the localized SSD profiles closely mirror those of the highly ordered nematic phase, albeit with significantly greater variance and more pronounced tails. However, the long-time MSD exhibits loss of the logarithmic super-diffusive corrections. This key departure signifies the structural choking of the gapless Goldstone modes, a direct consequence of 
a finite mass gap introduced by the fractured, multi-domain canted architecture. 
(iv) Role of Topological Defects: Finally, we demonstrate that a similar breakdown of super-diffusion occurs when topological defect lines (vortex strings) are introduced directly into the nematic phase. This underscores a overarching rule governing these systems: the presence of topological defects - whether discrete vortex lines or widespread interfacial domain walls - consistently disrupts long-range Goldstone fluctuations. Consequently, active super-diffusive behavior is systematically suppressed in any topologically frustrated or defective liquid crystalline environment.

The structural and transport characteristics of the ABP trajectories profoundly influence macroscopic processes, including targeted drug delivery, environmental dispersal, and cellular migration. In unfrustrated nematic liquid crystals, the swimming trajectory is strictly dictated by the coupling between hydrodynamic flows, particle geometry, and surface anchoring. This hydrodynamic coupling typically distinguishes two main classes of active matter: extensile (pusher) swimmers such as E. coli, flagellated bacteria, or catalytic Janus particles, which draw fluid inward laterally and expel it along their swimming axis to propel themselves parallel to ${\hat{\bf n}}$ ~\cite{goral2022frustrated}; and contractile (puller) swimmers such as Chlamydomonas algae or specialized phoretic particles which pull fluid along their symmetry axis and push it out transversely, moving predominantly perpendicular to ${\hat{\bf n}}$ ~\cite{Chi2020,goral2022frustrated}. These distinct operational modes emphasize the necessity of decoupling the parallel and perpendicular transport coefficients of the host medium. Our discrete GLL framework successfully accommodates this pusher-puller dichotomy by revealing how background microstructures alter these directional transport profiles.

Furthermore, the interplay of surface anchoring (planar vs. homeotropic) and particle morphology (rod-like vs. spherical) governs the local defect topology—generating characteristic configurations such as Saturn rings, hedgehogs, or boojums that exert elastic torques to bias the particle's heading relative to the director field~\cite{Mertelj2017}. While planar-anchored rods are elastically channeled along ${\hat{\bf n}}$ \cite{SelvinRobert2026}, spherical inclusions respond strictly according to local defect symmetry \cite{Lavrentovich2010}. Spatially patterned director fields and isolated topological defects (e.g., splay, bend, or $\pm1/2$ disclinations) generate localized elastic and hydrodynamic forces capable of steering active particles either along or transverse to the far-field director, providing a pathway for controlled transport within complex environments \cite{Lavrentovich2020}. Exploring how these distinct topological defects explicitly trap or guide active particles can be an interesting direction for future study.

While this work focused on the dilute limit of a single ABP where particle-particle interactions are negligible, the discrete modeling framework established here can be naturally extended to capture collective effects or evaluate other active classes, such as chiral or run-and-tumble swimmers. Additionally, structured liquid crystalline phases - including smectic layers, cholesteric helices, and bent-core phases - offer entirely new topological landscapes to manipulate and guide microscopic active transport. Ultimately, this study establishes a rigorous foundation for the systematic engineering of active motion across diverse activity regimes and complex liquid crystalline textures.

\subsection*{\bf Acknowledgments} RR acknowledges CSIR, India, for a research fellowship. RR, MA and VB gratefully acknowledge the HPC facility of IIT Delhi for computational resources. VB acknowledges ANRF, India, for a CORE research grant.

\appendix
\section{Symbols and parameters used in this study}

\renewcommand{\thetable}{\Alph{section}.\arabic{table}}
\setcounter{table}{0}
        
    \begin{table}[H]
        \begin{ruledtabular}
        \begin{tabular}{ll}
            \textbf{Symbol} & \textbf{Description} \\
            \hline
             $\mathbf{r}_a$& Position of the ABP \\
             \hline
             $\hat{\mathbf{p}}_a$& Orientation of the ABP \\
             \hline
             $\mathbf{r}_i$& Position of the $i^{th}$ nematogen \\
             \hline
             $\hat{\mathbf{n}}_i$& Orientation of the $i^{th}$ nematogen \\
             \hline
             $F_0$ & Magnitude of active force on the ABP\\
             \hline
             $a$ & Lattice constant\\
             \hline
             $\boldsymbol{\Omega}_a$ & Torque on the ABP due to its surroundings\\
             \hline
             $\boldsymbol{\Omega}_i$ & Torque on the $i^{th}$ nematogen due to its nn\\
             \hline
             $\boldsymbol{\Omega}_{i,a}$ & Torque on the $i^{th}$ nematogen due to ABP\\
             \hline
             $\boldsymbol{\Omega}_{an}$ & Torque due to thermal noise on the ABP\\
             \hline
             $\boldsymbol{\Omega}_{in}$ & Torque due to thermal noise on the $i^{th}$ nematogen \\
             \hline
             $\lambda$& The coefficient of the $P_4$ term in the GLL Hamiltonian\\
             \hline
             $\zeta_a$& Translational friction coefficient of ABP \\
             \hline
             $\xi_a$& Rotational friction coefficient of ABP \\
             \hline
             $\xi$& Rotational friction coefficient of nematogens \\
             \hline
             $\epsilon$& LC-LC interaction strength\\
             \hline
             $\epsilon_a$& ABP-LC interaction strength\\
             \hline
             $\sigma$& Decay length of the ABP-nematogen interaction\\
             
        \end{tabular}
        \end{ruledtabular}
        \captionsetup{justification=raggedright}
        \caption{List of symbols and their descriptions.}
        \label{Table A.1: model symbol list}
    \end{table}

    \newpage
    \begin{table}[H]
        \begin{ruledtabular}
        \begin{tabular}{ll}
             \textbf{Symbol} & \textbf{Description} \\
             \hline
             $T^* = k_BT/\epsilon$ & reduced bath temperature \\ 
             \hline
             $\tau = k_BTt/\xi$& Reduced time\\
             \hline
             $f_0= aF_0/k_BT$ & Reduced active force on the ABP\\
             \hline
             $\boldsymbol{\omega}_{\alpha} = \Omega_{\alpha}/k_BT$ & Reduced torque\\
             \hline
             $\tilde{\zeta}_a = a^2 \zeta_a / \xi$& Scaled translational friction coefficient of the ABP.\\
             \hline
             $\tilde{\xi}_a = \xi_a/ \xi$& Scaled rotation friction coefficient of the ABP.\\
        \end{tabular}
        \end{ruledtabular}
        \captionsetup{justification=raggedright}
        \caption{List of symbols and their descriptions used in the reduced (scaled) equations.}
        \label{Table A.2: reduced symbol list}
    \end{table}

    \begin{table}[H]
        \begin{ruledtabular}
        \begin{tabular}{ll}
             \textbf{Symbol} & \textbf{Description} \\
             \hline
             $a=1$& Lattice constant\\
             \hline
             $\epsilon=1$& LC-LC interaction strength\\
             \hline
             $\epsilon_a=1$& ABP-LC interaction strength\\
             \hline
             $f_0=5$ & Reduced active force on the ABP\\
             \hline
             $l=5$& Size-ratio of the ABP and nematogen\\
             \hline
             $\tilde{\zeta}_a = 3l/4$& Ratio of friction coefficients\\
             \hline
             $\tilde{\xi}_a = l^3$& Ratio of friction coefficients\\
             \hline
             $\sigma=2$& Decay length of the ABP-nematogen interaction\\
        \end{tabular}
        \end{ruledtabular}
        \captionsetup{justification=raggedright}
        \caption{Numerical values of different parameters used in the simulations.}
        \label{Table A.3: Symbols with values}
    \end{table}

    \begin{table}[H]
        \begin{ruledtabular}
        \begin{tabular}{lllllllll}
             \textbf{Parameters} & $a$&$\epsilon$&$\epsilon_a$&$f_0$&$l$&$\tilde{\zeta}_a$&$\tilde{\xi}_a$&$\sigma$ \\
             \hline
             \textbf{Values} & $1$&$1$&$1$&$5$&$5$&$3l/4$&$l^3$&$2a$ \\
             
        \end{tabular}
        \end{ruledtabular}
        \captionsetup{justification=raggedright}
        \caption{Numerical values of different parameters used in the simulations.}
        \label{Table A.4: Symbols with values}
    \end{table}

\newpage
\bibliographystyle{apsrev4-1}
\bibliography{ref}
\end{document}